\documentclass[]{aa}
\usepackage[english]{babel}
\usepackage{amsmath, amssymb, amsfonts}
\usepackage[]{graphicx}
\usepackage{natbib,twoopt}
\bibpunct{(}{)}{;}{a}{}{,} 
\usepackage{gensymb}

\title{Is the HD\,15115 inner disk really asymmetrical?\thanks{Based on data retrieved from the Gemini archive}} 

\titlerunning{Is the HD\,15115 inner disk really asymmetrical?}
\authorrunning{Mazoyer et al.}

\author{J.~Mazoyer\inst{1} \and A.~Boccaletti\inst{1} \and J.-C.~Augereau\inst{2,3} \and A.-M.~Lagrange\inst{2,3} \and R.~Galicher\inst{1} \and P.~Baudoz\inst{1}}

\institute{LESIA, Observatoire de Paris, CNRS, UPMC and Univ. Paris Diderot, 5 place Jules Janssen, 92190 Meudon, France. \email{johan.mazoyer@obspm.fr}\label{inst1} 
\and 
Univ. Grenoble Alpes, Institut de Plan\'e{}tologie et d\'{}Astrophysique (IPAG) F-38000 Grenoble, France
\and
CNRS, Institut de Plan\'e{}tologie et d\'{}Astrophysique (IPAG) F-38000 Grenoble, France
}


\abstract 
{Debris disks are intrinsically connected to the planetary system's formation and evolution. The development of high-contrast imaging techniques in the past 20 years is now allowing the detection of faint material around bright stars with high angular resolution, hence opening an avenue to study in detail the structures of circumstellar disks and their relation to planetary formation.}
{The purpose of this paper is to revisit the morphology of the almost edge-on debris disk around HD\,15115.}
{We analyzed data from the Gemini science archive obtained in 2009 and 2011 with the Near-Infrared Coronagraphic Imager instrument in the H and Ks bands using coronagraphy and angular differential imaging techniques. }
{We resolved the disk in both the H and Ks bands. We confirmed the position angles inferred by previous authors, as well as the brightness asymmetry, which is the origin of the object's nickname, the blue needle. We were able to detect the bow-like shape of the disk suspected from other observations. However, these new NICI images suggest the presence of a highly inclined ring-like disk of which we see the brighter side and the ansae located at 90 AU symmetrically about the star. The inner part is likely depleted of dust. The fainter side of the disk is suspected but not firmly detected, which also indicates a large anisotropic scattering factor.}
{The morphological symmetry of the disk contrasts with the obvious brightness asymmetry. This asymmetry may be explained by the coexistence of several types of grains in this disk and/or variable dust density. Interaction with the interstellar medium was invoked by previous authors as a possible explanation but other mechanisms may account for the brightness asymmetry, for instance a recent collision in the disk.}

\begin{document}
\maketitle

\section{Introduction}
\label{sec:intro}

Recently, several exoplanets were imaged around stars also harboring circumstellar dusty debris disks such as $\beta$ Pictoris \citep{Lagrange2009, Lagrange2010}, HR\,8799 \citep{Marois2008, Marois2010}, and HD\,95086 \citep{Rameau2013}. In the favorable case of $\beta$ Pic, simultaneous imaging of the disk and the planet \citep{Lagrange2012} confirmed the gravitational impact of the planet in the form of the well-known warp, as hypothesized earlier by \cite{Mouillet1997} and \cite{Augereau2001}. Studying the morphology of protoplanetary and debris disks is an essential tool for connecting peculiar structures (rings, gaps, warps, spirals) to the presence of planets although the determination of their orbital properties is not straightforward \citep{Krivov2010, MoroMartin2011}. 

Imaging of circumstellar disks in scattered light was for a long time limited to space telescopes. However, in the past 15 years, large ground-based telescopes combined with adaptive optics systems, stellar coronagraphs, and differential imaging techniques have allowed the study of the morphology of a handful of protoplanetary and debris disks. One of these is the disk surrounding HD~15115, an F2 star located at $45.2\pm 1.3$\,pc \citep{VanLeeuwen2007}. The age of HD~15115 is estimated between 12 Myr and 100 Myr, so rather uncertain. Using ISO data, \cite{Silverstone2000} showed that HD~15115 exhibits strong infrared excess emission likely due to a dusty debris disk.

The first resolved images of the HD~15115 were published in \cite{Kalas2007} using data from HST/ACS in the visible and from NIRC2/Keck in H band, both using coronagraphs. The disk appeared nearly edge-on and oriented east-west at a position angle (PA) of  $278.5\degree\pm0.5\degree$, measured from the west side.
These observations revealed a strong east-west brightness asymmetry (the west side appears brighter and more extended), which together with its color, gave its name to the object: the blue needle. \citet{Kalas2007} attempted to explain the disk asymmetry by the presence of a perturbing star in the system (HIP\,12545). \citet{Debes2008} used HST/NICMOS data to confirm the PA of the disk to be $278.9\degree \pm 0.2\degree$ for the outer part of the west side ($> 1.7$''), and noted a slight difference with the east side PA which is $276.6\degree \pm 1.3\degree$, although it seems marginal given the error bars.
In addition, at shorter separations ($< 1.7$''), they found the PA to increase inversely with the distance to the star.  \citet{Debes2008} considered a warp or a tilted disk to be a plausible explanation, but did not observe an opposite trend in the east side to support this hypothesis.
They also refuted the possibility of HIP\,12545 being a perturber, backtracking the previous positions of the two stars. \citet{Rodigas2012} analyzed data from LBT/PISCES in Ks and LBTI/LMIR in L'. 
While the east-west asymmetry is still visible in Ks, the disk appears almost symmetrical and is gray from K to L, contrary to the blue color measured at shorter wavelengths. 
Using forward modeling, they found the disk to be bowed, which definitely rules out the warp assumption. Importantly, \citet{Rodigas2012} claimed that their analysis supports a gap at a separation smaller than 1", in apparent agreement with the SED modeling by \citet{Moor2011}. Finally, \citet{Schneider2014} reported the latest results of the HST/STIS GO/12228 program. They confirm disk bowing as close as 0.4'', likely attributed to the brightest part of a disk in a nearly edge-on geometry, which also implies a partial inner clearing.

\begin{table*}[ht]
\caption{Log of the observations \label{table:observations}}
\begin{center}
\begin{tabular}{lcccccc}
 \hline \hline
\vspace{-0.15cm} \\
Date \vspace{0.05cm}  & Filters & Integration time & Coadds & Number of & Total integration & Total parall.\\ 
						& 		&  per coadd [s]	& per frames & frames 	& time [s] 			& angle rotation\vspace{0.1cm} \\
\hline
 \vspace{-0.15cm} \\
2009 Dec 4  & CH4S / CH4L $^{\text{a}}$ 	& 20.14 & 3 & 43 & 2598.06 & 17. 42$\degree$ \\
2011 Nov 7  & Ks / H $^{\text{b}}$		& 4.94 	& 12 & 20 & 1185.6 & 8.57$\degree$ \\
2011 Nov 22 & Ks / H $^{\text{b}}$		& 9.88 	& 6 & 40 & 2371.2 & 13.71$\degree$ \vspace{0.1cm} \\ 
 \hline
\vspace{-0.5cm}
\end{tabular}
 \end{center}
 \tablefoot{ \small 
$^{\text{a}}$ Narrow band filters in H band. CH4S: $\lambda = 1.652$ $\mu$m and $\Delta \lambda = 0.066$ $\mu$m. CH4L: $\lambda = 1.578$ $\mu$m and $\Delta \lambda = 0.062$ $\mu$m. \\
$^{\text{b}}$ Ks: $\lambda = 2.15$ $\mu$m and $\Delta \lambda = 0.16$ $\mu$m. H: $\lambda = 1.65$ $\mu$m and $\Delta \lambda = 0.29$ $\mu$m.}
\end{table*}

This paper presents a new analysis of the HD\,15115 disk images obtained at Gemini with NICI. In Sect.~\ref{sec:obs} we describe the observations and data reduction. Sect.~\ref{sec:morpho} reports on the morphological measurements performed in the disk image. We present our results of forward modeling in Sect.~\ref{sec:fwmodel}, 
and photometric measurements are presented in Sect.~\ref{sec:photom}. Results are summarized in Sect.~\ref{sec:ccl}.

\section{Observations and data reduction}
\label{sec:obs}

We used the archival data from the near-infrared coronagraphic imager \citep[NICI; ][]{Toomey2003} installed on the Gemini South telescope, now decommissioned. NICI is a near-IR (1-5$\mu$m) coronagraphic instrument that can be operated in pupil-tracking mode and that uses two channels, separated with a beam splitter, to simultaneously image the same field with two detectors. Each channel has its own set of near-IR filters. This design, 
conceived to search for young exoplanets and brown dwarfs, allows the data to be simultaneously processed with angular differential imaging \citep[ADI, ][]{Marois2006} and spectral differential imaging \citep[SSDI, ][]{Racine1999}. The coronagraphic masks are semi-transparent Lyot occulters of various radius (0.22" to 0.90"). NICI has already successfully detected disks around HD\,142527 \citep{Casassus2013} and around HD\,100546 \citep{Boccaletti2013b}, for instance. 

\begin{figure*}
\begin{center}
\centerline{
\includegraphics[width=0.495\textwidth]{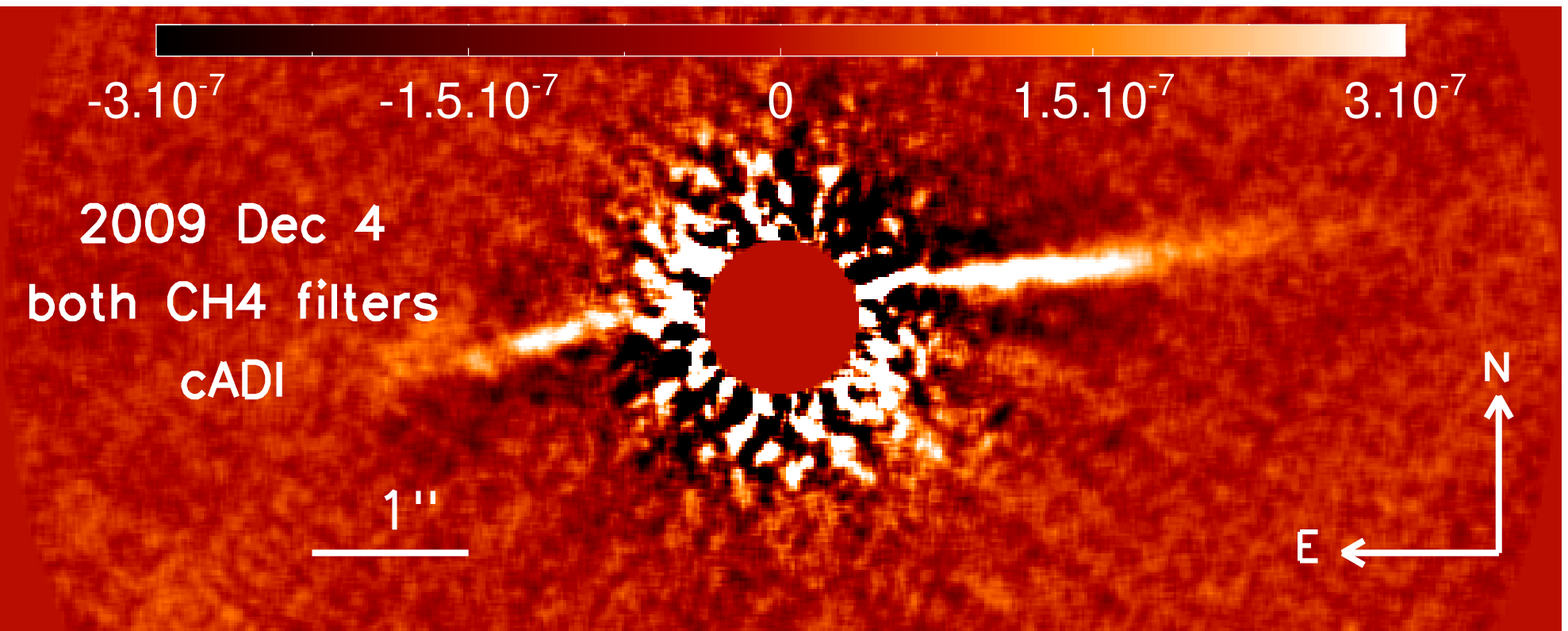}
\includegraphics[width=0.495\textwidth]{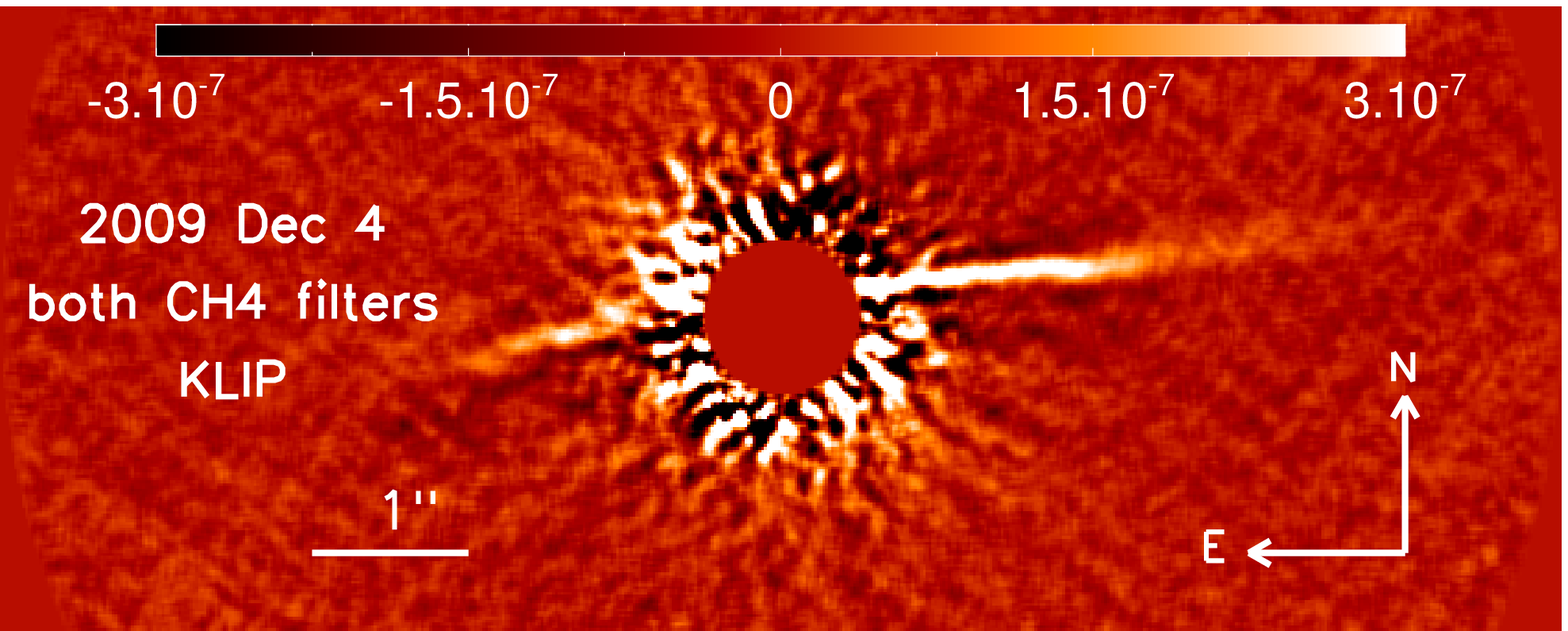}}
\centerline{
\includegraphics[width=0.495\textwidth]{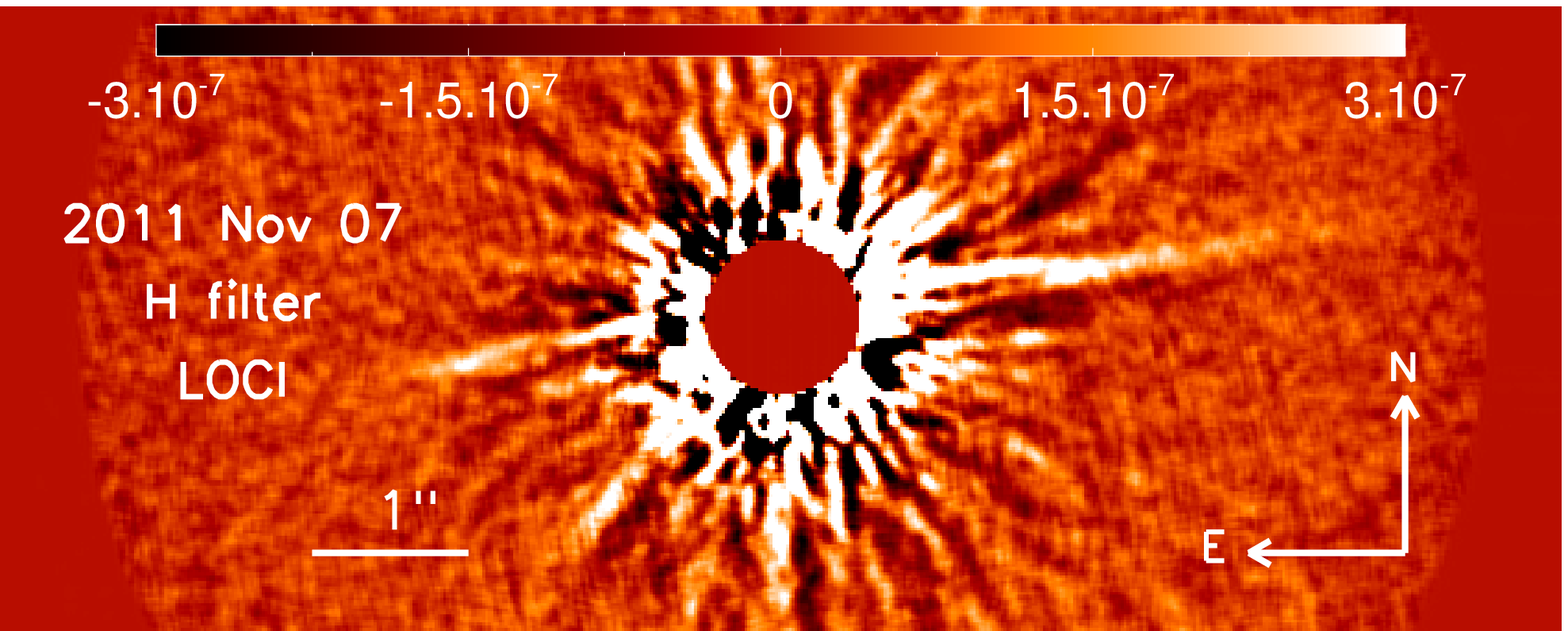} 
\includegraphics[width=0.495\textwidth]{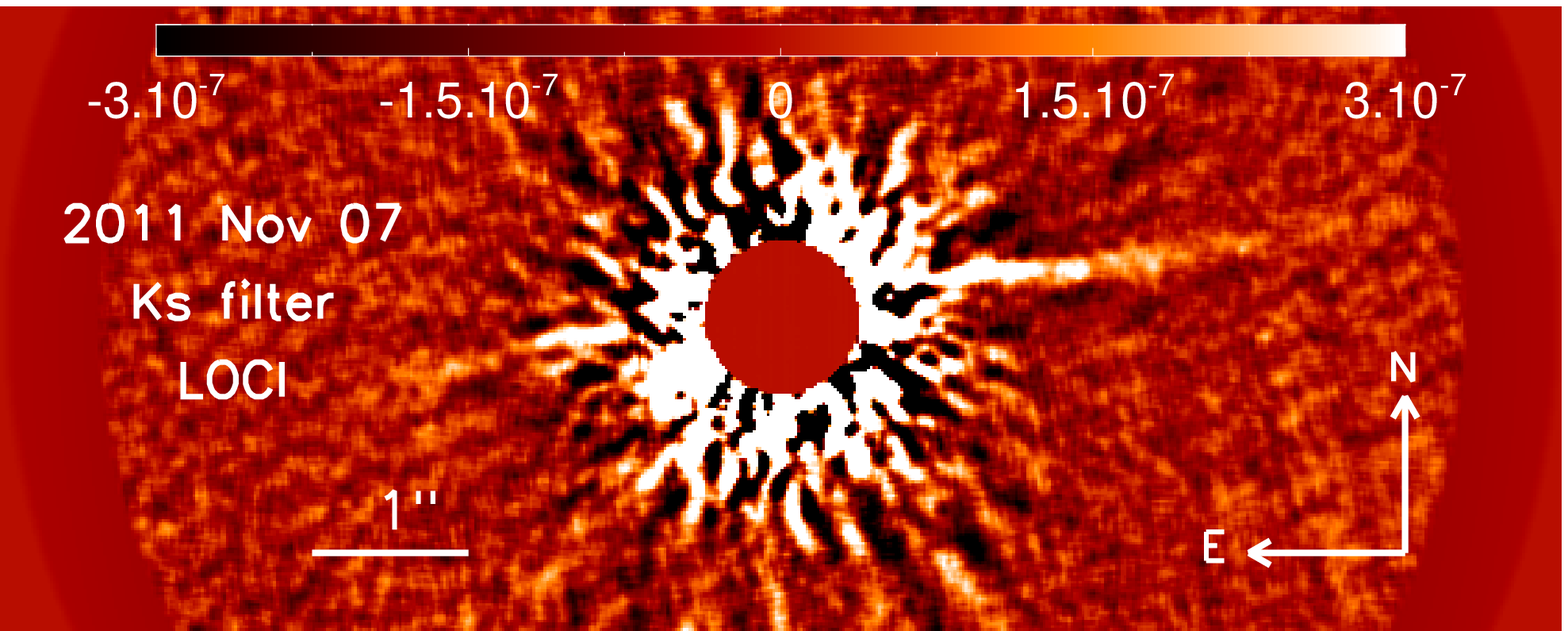}}
\centerline{
\includegraphics[width=0.495\textwidth]{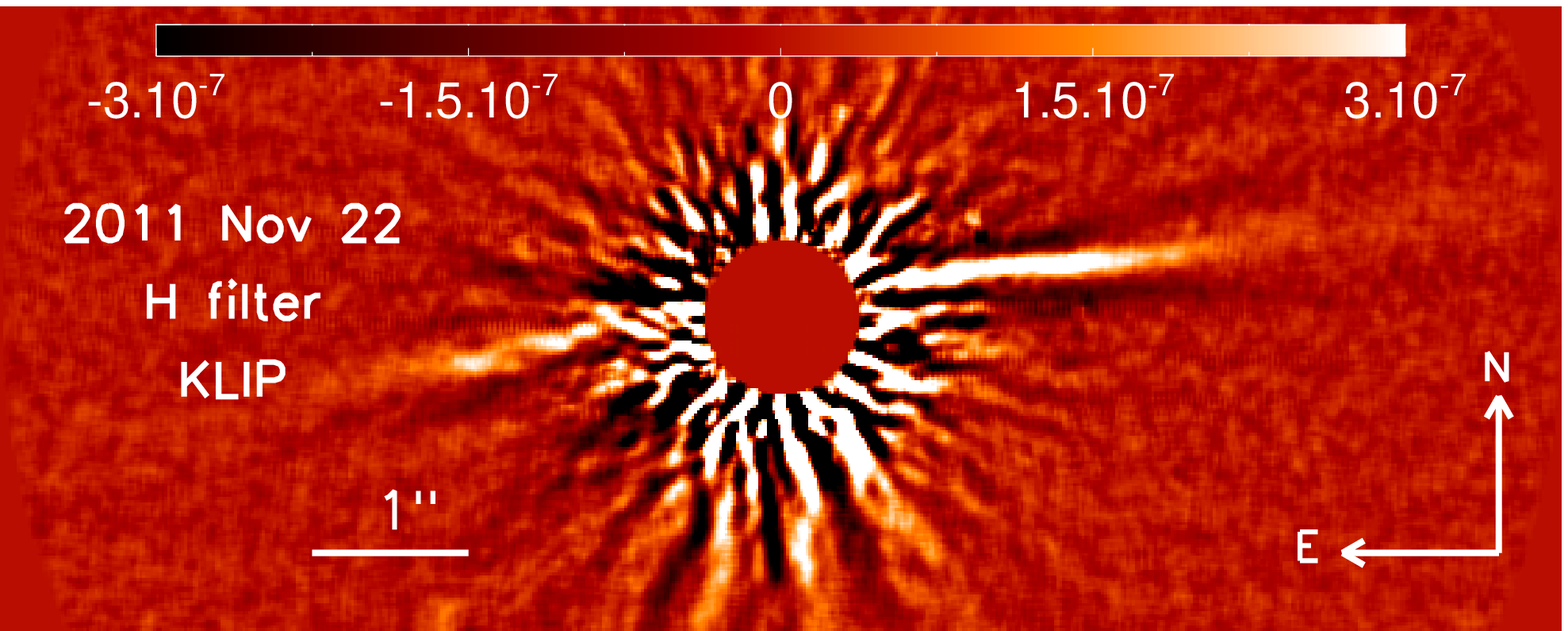} 
\includegraphics[width=0.495\textwidth]{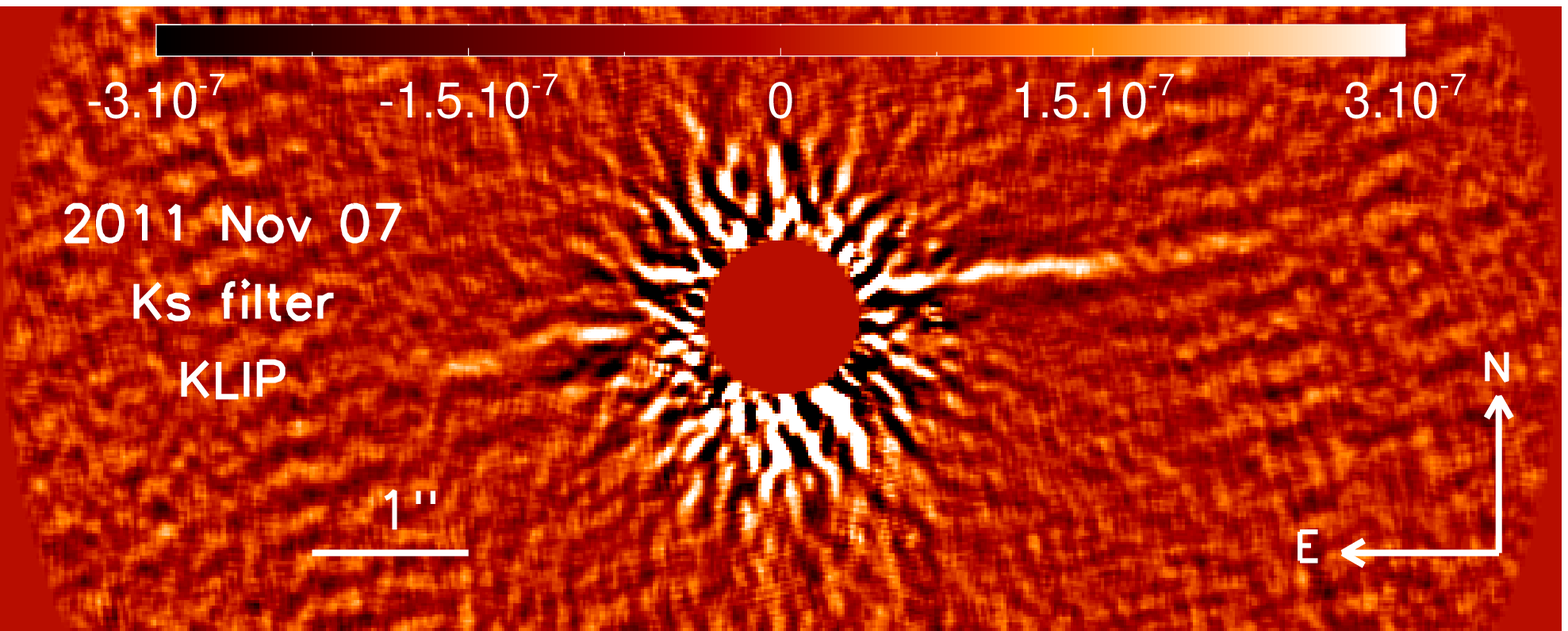}}
\end{center}
\caption[ADI_disk]
{\label{fig:ADI_disk} Various cADI, LOCI, and KLIP images of the disk around HD~15115. Top : cADI and KLIP images of the 2009 December 4 data (CH4S and CH4L filters combined). Middle : LOCI images in H and Ks band of the 2011 November 7 data. Bottom, left : KLIP image in H band of the 2011 November 22 data. Bottom, right : KLIP image in Ks band of the 2011 November 07 data. The images are smoothed at a scale of 1.5 resolution element. Color scales indicate the contrast with respect to the maximum intensity in the PSF image.}
\end{figure*}

We analyzed three sets of data as part of the NICI planet-finding campaign (program GS-2011B-Q-500, PI: M. Liu) obtained in December 2009 and November 2011 as described in Table~\ref{table:observations}. For each observation we indicate the date (Col. 1), the filters used in each channel (Col. 2) and the integration time of each of the 12 coadds (Col. 3). The number of frames (Col. 4) and the total integration time given in Col. 5 account for the frame selection in the data reduction. Lastly, the amplitude of the parallactic angle variation during the sequence is indicated in Col. 6. For all of these observation, the radius of the focal plane occulting mask was $0.32''$.

We followed the data reduction described in \citet{Boccaletti2013a}. Each single frame is dark subtracted, then flat field and bad pixel corrected. The dark images were sometimes unavailable for the specific integration time we used. In that case we used the dark images with the closest integration time and numerically corrected the weft visible in some images. The registration of frames is achieved in two steps, first a coarse alignment at one pixel accuracy, then in a second step, the central star leakage through the semi-transparent mask provides the necessary signal for measuring the star position more precisely. \citet{Boccaletti2013b} has demonstrated a registration accuracy of 0.2 pixel (1 pixel $\simeq$ 17.9 mas). We assume the same level of performance is met in the present case because the procedure is identical and the data is of similar quality. In the cases where the central coronagraphic spot is saturated, we used a cross correlation of frames measured in the unsaturated area surrounding the central peak. We decided to remove from the cube the images where the mask is not fully centered on the star.

There are no non-coronagraphic unsaturated images of this star recorded the same night. Therefore, we built the star point spread function (PSF) from the central unsaturated coronagraphic spot and used the transmission values found by \citet{Wahhaj2011} for the 0.32'' semi-transparent mask ($5.94 \pm 0.02$ mag and $5.70 \pm 0.03$ mag in the H and Ks bands). 
These PSFs are used to normalize the photometry of the images and obtain contrast levels. The final product is a temporal data cube of registered frames normalized to the PSF maximum intensity. The PSF FWHM measured in the central spot is 60 mas.

These data were processed with different algorithms to enhance the contrast. First of all, we used the classical ADI method \citep[cADI, ][]{Marois2006} on these images. However, because of the very small number of images and angle rotation in the last two sets of data (in 2011), cADI treatment produced poor results. 
The LOCI algorithm \citep[locally optimized combination of image; ][]{Lafreniere2007} was also applied. The efficiency of this algorithm has been repeatedly proven for high-contrast imaging of circumstellar disks \citep{Thalmann2010, Lagrange2012, Rodigas2012, Currie2012, Boccaletti2013b}. The LOCI parameters are tuned for two purposes simultaneously: an extended object and small field rotation. 
For the first purpose, we used a large optimization area. We took $N_A$ \citep[size of the optimization zone in PSF footprints; ][]{Lafreniere2007} of 3,000 PSF footprints, as in the conservative approach proposed by \citet{Thalmann2010}. The second one was mitigated with a rather small frame selection criteria : \cite{Lafreniere2007} defines the $N_\delta$ parameter as the minimum radial movement of the image of a source point between two consecutive selected frames. We chose $N_\delta$ = 0.75 x FWHM. In addition, the radial width of the subtraction zone $d_r$ was set to 3 x FWHM and the ratio of the radial to azimuthal width was set to 1.

Finally, we used the KLIP algorithm \citep[Karhunen-Lo{\`e}ve Image Projection, ][]{Soummer2012}. This method uses a principal component analysis where the science image is projected onto an orthogonal basis, named Karhunen-Lo{\`e}ve (KL) basis. The $n$ first KL vectors are retained to build a reference image hence subtracted to each frame of the data cube. This method is more suited to forward modeling (see Sect.~\ref{sec:model}) than LOCI, since the KL basis is independent of the science target.
Moreover, this PSF subtraction technique has already been used for disk analysis: HD\,100546 \citep{Boccaletti2013b}, HIP\,79977 \citep{Thalmann2013}, and HD\,32297 \citep{Rodigas2014}. Once again, the large spatial extension of the considered astronomical object requires a special parameterization and we limit the analysis to the first KL vectors ($n < 5$).

LOCI and KLIP were applied on the data cubes for the three epochs separately: December 2009 (CH4S/CH4S), and November 7 and 22, 2011 (Ks and H). Since we had no intention of performing spectral differential imaging in CH4 bands we summed the two channels from 2009 to improve the signal-to-noise ratio. 

\section{Results and morphological analysis}
\label{sec:morpho}

\subsection{The resulting images}
\label{sec:pres}

A selection of these images is presented in Fig. \ref{fig:ADI_disk} for various epochs, bands, and processing techniques. 
The disk clearly appears in all images nearly aligned with the east and west direction.
The east-west brightness asymmetry previously reported  \citep{Kalas2007, Debes2008, Rodigas2012} is an obvious feature of these images. We did not observe the inversion of the disk orientation in the east side inward to 1.5'', as seen in \citet{Rodigas2012}. On the contrary, the maximum intensity of the disk is clearly located above a median plane consistent with a fully bowed geometry. It is plausible that this former feature was an artifact. 
In addition, the disk in Ks seems fainter than in the H band. 

\begin{figure*}
\begin{center}
\includegraphics[width=0.99\textwidth]{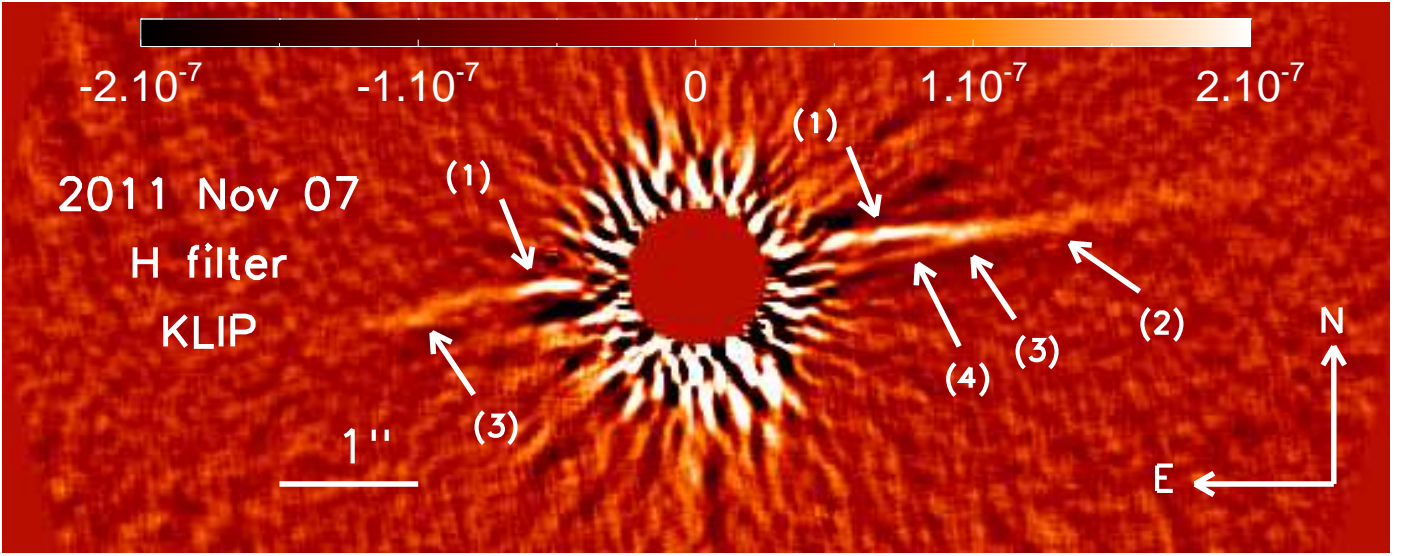}
\end{center}
\caption[ADI_disk]
{\label{fig:ADI_disk_big} Final KLIP image of the disk around HD~15115 in H band on November 7, 2011. We have indicated the zones of interest. The image is smoothed at a scale of 1.5 resolution element. The color scale indicates the contrast with respect to the maximum intensity in the PSF image.}
\end{figure*}

The analysis of contrasts in raw coronagraphic images led us to conclude that the November 7, 2011 sequence was of better quality (in terms of AO correction and stability and for separations larger than 1") than the others, despite a smaller field rotation. 
A KLIP processed image is shown in large format in Fig.~\ref{fig:ADI_disk_big}. We note that the disk image can be divided into two regions:
\begin{itemize}
\item the inner part ($\lesssim$2"), labeled (1) in Fig. \ref{fig:ADI_disk_big} 
on each side of the star, which definitely confirms the bow-like shape first observed by \citet{Rodigas2012} and \citet{Schneider2014};
\item the outer part ($\gtrsim$2''), labeled (2) in Fig. \ref{fig:ADI_disk_big}. This component, more visible in the west part, defines the global orientation of the disk, and confers to the disk its needle shape.
\end{itemize}

More importantly, these NICI data are the first to suggest a ring-like inner disk of which we see the upper brightest part. This might imply  dust depletion within a radius of about 2". In the November 7, 2011 data, the ansae of this ring (labeled (3) in Fig. \ref{fig:ADI_disk_big}) are visible on the two sides, while the November 22, 2011 data (Fig. \ref{fig:ADI_disk}, bottom left) shows this feature only on the east side. Finally, the bottom part of the ring, below the midplane might be visible in the western part in Fig. \ref{fig:ADI_disk_big} (labeled (4)).

A simple ellipse fitting of the upper part of the suspected ring, directly in a KLIP image from Nov 7, 2011, yields a first rough estimation of the disk parameter. We found $\text{PA}\approx 98.5\degree$, $i\approx 86.2\degree$, $\text{radius}\approx 2.0"$. 
Noticeably, the ring appears almost symmetrical in size, which contrasts with the brightness asymmetry. In addition, we did not measure any significant offset along the major axis nor along the minor axis with respect to the star's position.

\subsection{Signal-to-noise ratio}
\label{sec:snr}

\begin{figure*}[ht!]
\centerline{\includegraphics[width=0.495\textwidth]{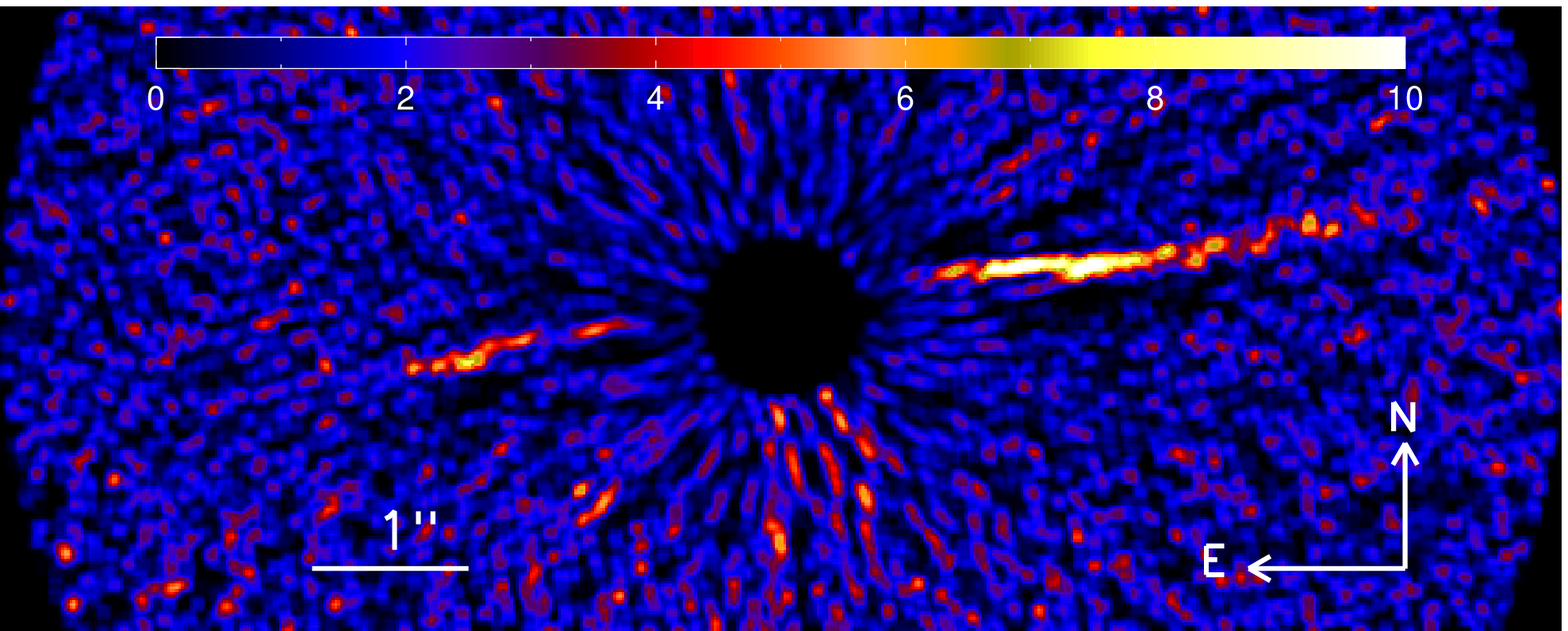}
\includegraphics[width=0.495\textwidth]{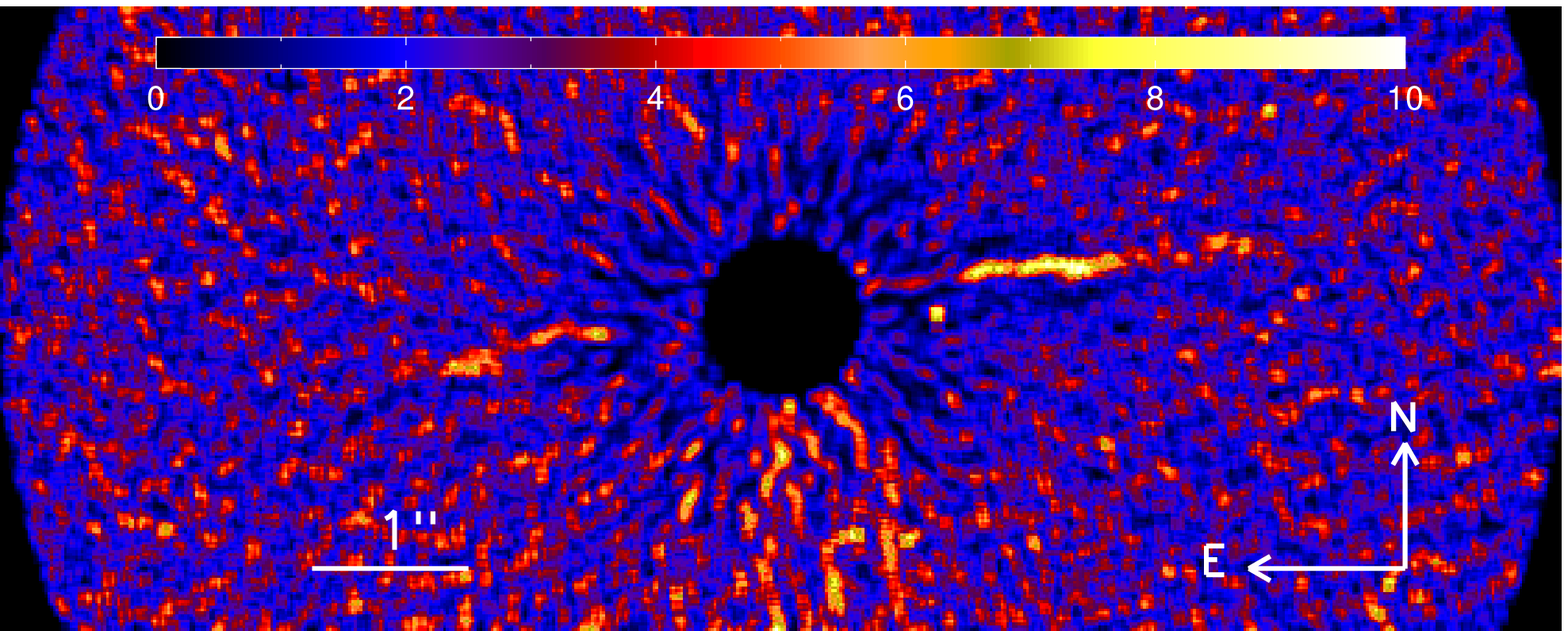}}
\caption[]{
\label{fig:SNR}signal-to-noise ratio maps, calculated for the H (left) and Ks (right) bands in the November 7, 2011 data.  The images are smoothed at a scale of 1.5 resolution element. Color scales indicate the S/N per resolution element.}
\end{figure*}

We calculated the signal-to-noise ratio (S/N) in the H (Fig. \ref{fig:SNR}, left) and Ks (Fig. \ref{fig:SNR}, right) bands in the November 7, 2011 data. 
The noise is estimated as the azimuthal standard deviation in the image smoothed at one resolution element. The S/N maps in Fig. \ref{fig:SNR} are the original images divided by the noise maps.
The disk is detected as close as a separation of 1" on both sides. In the Ks band the disk has a lower S/N than in the H band as we can expect from its blue color \citep{Debes2008}. Finally, the S/N map confirms the brightness asymmetry, and the rather symmetrical ring-like pattern of the inner disk.

The following subsections are intended to test the reliability of the ring-like feature. For this purpose, we measured precisely the morphological characteristics of the inner disk: position angle, inclination and the trace of the disk. First, we start by introducing the model we will be using to determine some of the disk properties.

\subsection{Model}
\label{sec:model}
The GRaTer code to model debris disks is presented in \citet{Augereau1999} and \citet{Lebreton2012}. GRaTer calculates synthetic scattered-light images of optically thin disks, assuming axial symmetry about their rotation axis. 
The model considers the geometry of a cold debris disk: a dust belt located at the angular distance $r_0$, with a density radially decreasing inward and outward as power laws with slopes $\alpha_{in}$ and $\alpha_{out}$. In this case, while assuming a Gaussian vertical profile and a scale-height proportional to the distance to the star, the dust number density is given by (in cylindrical coordinates)

\begin{equation}
\label{eq:model_grater}
\resizebox{.9\hsize}{!} {$n(r,z) = n_0
\sqrt{2}\left[\left(\frac{r}{r_0}\right)^{-2\alpha_{in}} +
  \left(\frac{r}{r_0}\right)^{-2\alpha_{out}}\right]^{-\frac{1}{2}}
\exp{\left(\left(\frac{-|z|}{H_0\left(\frac{r}{r_0}\right)}\right)^{2}\right)}$}
\end{equation}
where $n_0$ and $H_0$ are the midplane number-density and the disk scale-height at the reference distance to the star $r=r_0$. We also define $h= H_0/r_0$, the disk aspect ratio at $r = r_0$. A ray-traced image was then calculated assuming a disk inclination $i$ from pole-on viewing, and anisotropic scattering modeled by an \citet{Henyey1941} phase function with asymmetry parameter $g$ between $-1$ (pure backward scattering) and $1$ (forward scattering). 

This model was used intensively on several occasions for the analysis of high-contrast observations of disks, \citep{Lagrange2012, Boccaletti2012, Milli2012, Milli2014} to overcome the photometric and astrometric biases introduced by ADI in disk images. Since it is difficult to reverse the problem in a simple way simulated disks can instead be implanted in the temporal data cube and processed simultaneously with the real disk. 

\subsection{Position angle measurements}
\label{sec:PA}

To compare with previous studies of HD\,15115 \citep{Kalas2007,Debes2008,Rodigas2012}, we first measured the position angle (PA) of the disk. We used the strategy developed for very inclined disks presented in \citet{Lagrange2012} and \citet{Boccaletti2012}. 
Assuming a starting PA, $\text{PA}_0 = 98\degree$, the image is rotated by $\text{PA}_0$ to roughly align the disk with the horizontal axis.
Then, a Gaussian profile is fitted on each column to determine the position of maximum intensity as a function of separation from the star, which refers to the trace of the disk. We measured the PA between $2''$ and $3''$ in the part of the disk outside the bowed area, but inside a region where the disk has a sufficient S/N. A trade of analysis led us to select four close regions: $[2.00'',2.90'']$, $[2.06'',2.95'']$, $[2.09'',2.99'']$, and $[2.14'',3.04'']$ on both the east and west sides, to mitigate local brightness variations. 
In these four regions the disk profile is fitted with a linear function (elevation to the midplane versus radial separation), and the averaged values give the local slope of the disk.
The procedure is repeated iteratively in a range of orientations, $\text{PA}_0 \pm 2\degree$  with steps of 0.01$\degree$. We define the PA as the orientation that minimizes this slope in the considered regions, for the east and west sides separately.
To assess the error bar of our method, it is repeated on a set of LOCI and KLIP processed images for the three epochs (ten images in total). We obtained $\text{PA}=98.1 \degree\pm0.7\degree$ on the east side and $\text{PA}=(99.5+180)\degree\pm0.4\degree$ on the west side, which agrees well with previous observations. Therefore, the average east-west PA is $98.8 \degree\pm0.4\degree$ and the east-west PA difference  is $1.4 \degree\pm0.8\degree$. \citet{Debes2008} also reported a difference of orientation between the two sides of about $2\degree$, while it was not a feasible measurement in the \citet{Rodigas2012} image. 

\begin{figure}
\begin{center}
\begin{tabular}{c}
\includegraphics[trim=2cm 2.35cm 1.2cm 1cm, clip=true,width=0.45\textwidth]{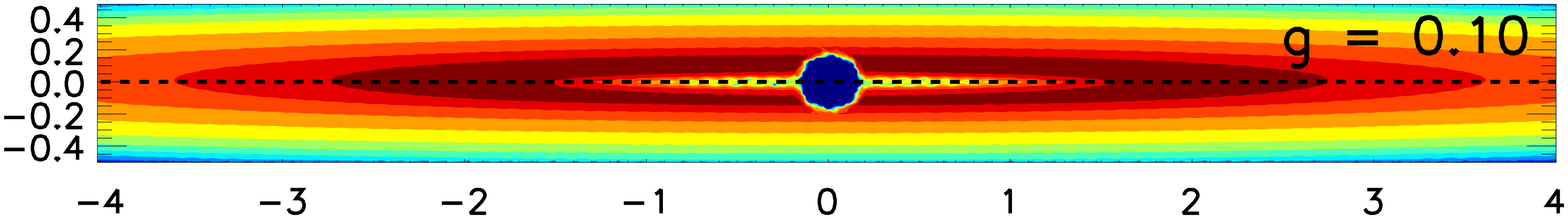}\\
\includegraphics[trim=2cm 2.35cm 1.2cm 1cm, clip=true,width=0.45\textwidth]{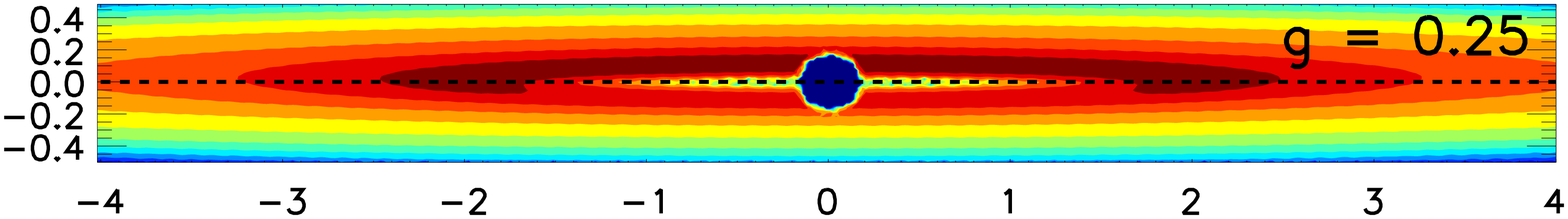}\\
\includegraphics[trim=2cm 2.35cm 1.2cm 1cm, clip=true,width=0.45\textwidth]{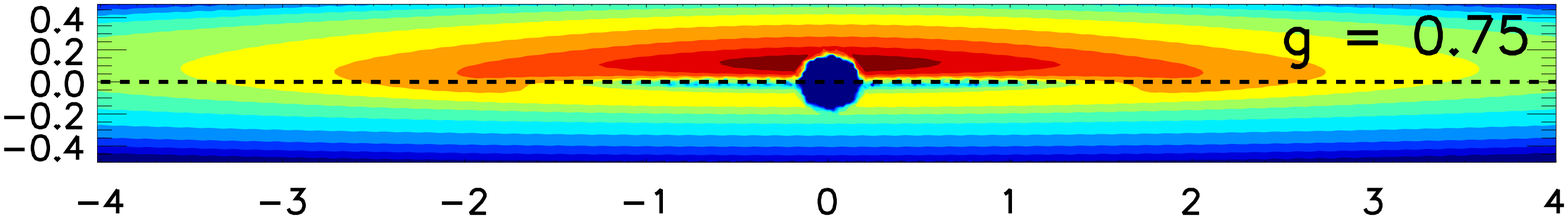}\\
\includegraphics[trim=1.65cm 0.5cm 0.85cm 0.7cm, clip=true,width=0.45\textwidth]{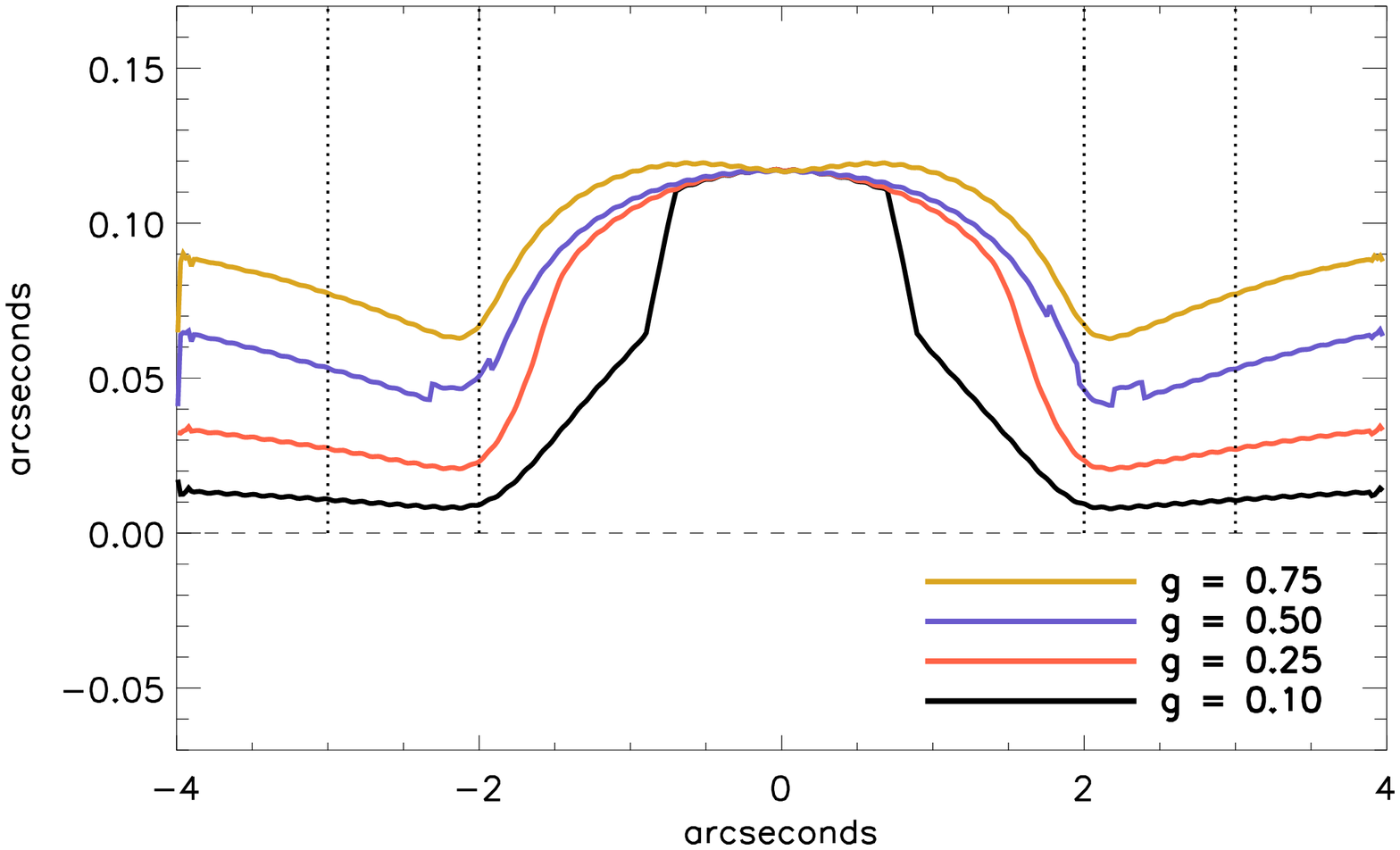}
\end{tabular}
\end{center}
\caption[spine_models]
{\label{fig:spine_models} Models of a very inclined disk ($i=87\degree$) with a ring at $2''=90\,AU$ generated for three anisotropic scattering factors ($|g|$ = 0.10, 0.25, and 0.75 from top to bottom). The bottom panel shows the trace of the disk (elevation in arcseconds with respect to the position of the midplane) measured for each model (black, red, and blue lines). The measured elevation is perturbed for the $|g| = 0.10$ disk between 1'' and 2'', since we are fitting a single Gaussian profile while two appear on each part of the midplane as long as the disk image is nearly symmetrical. We indicate with black vertical dotted line the zones where we measure the slope and the elevation of the trace.}
\end{figure}

The difference of PA between the two sides of the disk and beyond the peak position of the ring, 
is actually inherent to a disk with a nonzero anisotropic scattering factor ($|g|$). We demonstrate this effect by producing disk simulations with all parameters of Eq.~\ref{eq:model_grater} fixed ($\text{PA}=90\degree$,  $i=87\degree$, $r_0 = 2''$, $\alpha_{in}=10$, $\alpha_{out}=-4$), except for $|g|$ which we successively set to 0.10, 0.25, 0.50, and 0.75. Figure \ref{fig:spine_models} shows the corresponding images for some of these models. The disk intensity tends to distribute above the midplane while $|g|$ increases. We plot in the same figure (bottom) the traces of these disks retrieved with our method. We indicate with black vertical dotted lines the zones where we measure the slope and the elevation of the disk ($[2'',3'']$). In the outer part of the disk, beyond the ring located at 2", the elevation of the disk with respect to the midplane shows two trends: both the slope and the elevation increase with $|g|$. As a result, the two sides of the disk are no longer aligned (the slopes are of opposite signs) and no longer intersect the star. Therefore, measuring such a behavior could be a way to estimate $|g|$ once other disk parameters are set.

This study on simulted disks showed that 
\begin{itemize}
\item the mean of the values of the two slopes is equal to the PA ($90\degree$ in the case of Fig.~\ref{fig:spine_models});
\item all other parameters fixed, the difference of slopes and the elevation from the midplane go linearly with the $|g|$ parameter, from which we will estimate the $|g|$ parameter of our disk in the next sections.
\end{itemize}

The same conclusions were found if the analysis is applied on models processed with KLIP (see Sect.~\ref{sec:fwmodel}). We deduced that the PA of the disk is the average of the east and west, and is $98.8\degree\pm0.4\degree$, in good agreement with the most accurate measurement to date \citep[99.1$\degree$ in ][]{Schneider2014}. 

\subsection{Trace of the disk}
\label{sec:spine}

\begin{figure}[t]
\centerline{
\includegraphics[width=9cm]{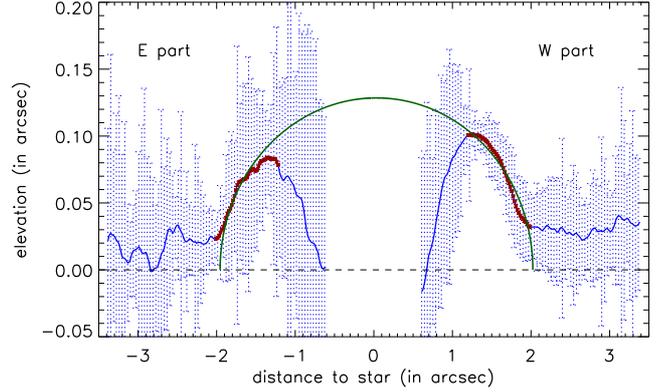}}
\caption{Trace of the disk measured in several images (LOCI and KLIP, three epochs, two filters). The solid blue line shows the averaged trace of the disk, and the dashed lines stand for the dispersion (maximum and minimum values). We selected the inner disk values ($<2''$) where S/N$>1$ (plotted in red dots in the figure), and used them to fit an ellipse (green line).}
\label{fig:hd15115spine}
\end{figure}

We now consider the LOCI and KLIP images for all epochs and all bands (a total of ten images) to measure the trace of the disk. Assuming the disk PA derived in Sect.~\ref{sec:PA}, the images are derotated to orient the disk horizontally. 
The location of maximum intensity along the midplane is measured as in the previous section in a range of separation of [-3.5", 3.5"]. Figure~\ref{fig:hd15115spine} shows the averaged trace for all images together with the dispersion (maximum and minimum values) among these images. The regions beyond 2" appears nearly horizontal as expected, but we find a non-zero elevation ($\sim0.02\pm0.01"$ to the east and $\sim0.03\pm0.01"$ to the west) toward the north, likely in agreement with the analysis performed in Fig.~\ref{fig:spine_models}. Therefore, an offset of the disk as proposed in \citet{Rodigas2012} is not necessarily responsible for this characteristic as the observed elevation could be explained solely with anisotropic scattering. 

\begin{figure}[t]
\centerline{
\includegraphics[width=9cm]{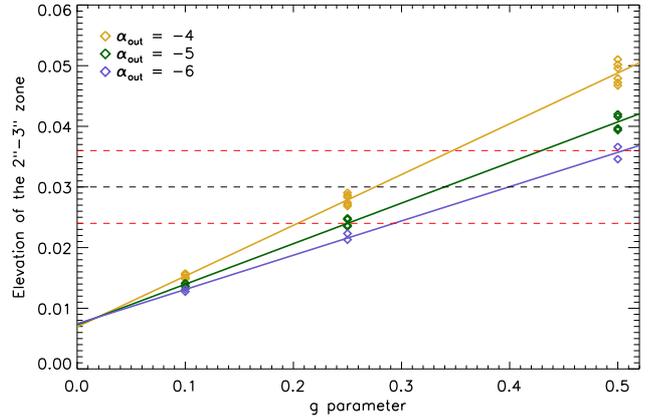}}
\caption{Elevation from the midplane in the 2-3" region versus the anisotropic scattering factor $|g|$ as measured in the model images (diamonds) for various model parameters. The solid lines are linear fits to the data points. The dashed lines give the measured value (black) of the elevation in the HD\,15115 disk image together with the dispersion (red).}
\label{fig:elevation}
\end{figure}

As proposed in Sect.~\ref{sec:PA} we compared the elevation measured in the data to the elevation measured in the models versus the $|g|$ parameter. Figure \ref{fig:elevation} shows the linear relationship between the elevation and $|g|$, determined on simulated disks. We fixed $r_0 = 2''$, $i = 87\degree$ and plotted this elevation measured in the 2 to 3" region for $\alpha_{out}=$ -4, -5, -6; $\alpha_{in}=$ 2, 5, 10; and $h=$ 0.01, 0.02. Taking into account other model parameters and the uncertainty on the elevation, the likely range of $|g|$ beyond $2''$ is comprised between 0.2 and 0.5, corresponding to an elevation of $\sim0.03\pm0.01"$.

In Fig.~\ref{fig:hd15115spine}, the inner region ($<2"$) unambiguously reveals the bow-like shape. Using the points between 1'' and 2 '' (marked with red dots), we fit an ellipse which allows us to determine the morphology of the corresponding suspected ring. We set the orientation to $0\degree$ since the disk image is already horizontal in this plot with the PA measured in the previous section, and the position of the ellipse along the minor axis to be located onto the star (no offset along this axis). 
The star position is known with a very good accuracy since the coronagraph is semi-transparent. The best-fit ellipse is plotted in green. Its parameters are a semi-major axis of 1.99", a semi-minor axis of 0.13", and a very small offset of 0.04" along the midplane to the west, which is mostly consistent with a ring centered onto the star. From that, we concluded that the disk is mostly geometrically symmetrical about the star, which contrasts with the somewhat pronounced brightness asymmetry. In addition, at a distance of 45.2\,pc, our measurement corresponds to a ring located at $\sim90$\,AU, a value which is off by a factor of 2 with respect to the SED fitting based on a blackbody assumption \citep{Moor2011}. This factor is within the range observed for other debris disks. Finally, the ellipse fitting yields an inclination of $86.2\degree$ about the line of sight, in total agreement with \cite{Rodigas2012} and our first coarse estimation.

To summarize, we measured PA=$98.8\degree\pm0.4\degree$ and $i=86.2\degree$. These values are in complete agreement with previous studies. We came to the conclusion that the disk is likely a ring-like disk with a radius of $1.99'' \simeq 90$\,AU, and shows practically no offset and no asymmetry in shape, while it shows a strong brightness asymmetry. We were finally able to give an estimation of the $|g|$ parameter between 0.2 and 0.5 beyond the peak brightness position at ~$2''$ (90 AU).

\section{Forward modeling}
\label{sec:fwmodel}

\begin{figure}[t]
\centerline{
\includegraphics[width=9cm]{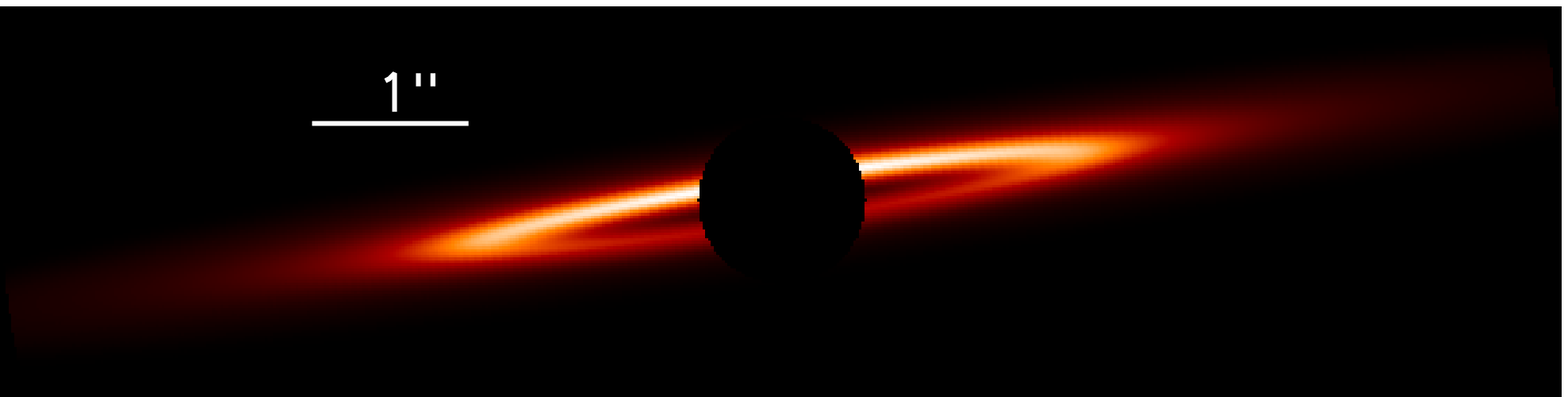}}
\centerline{
\includegraphics[width=9cm]{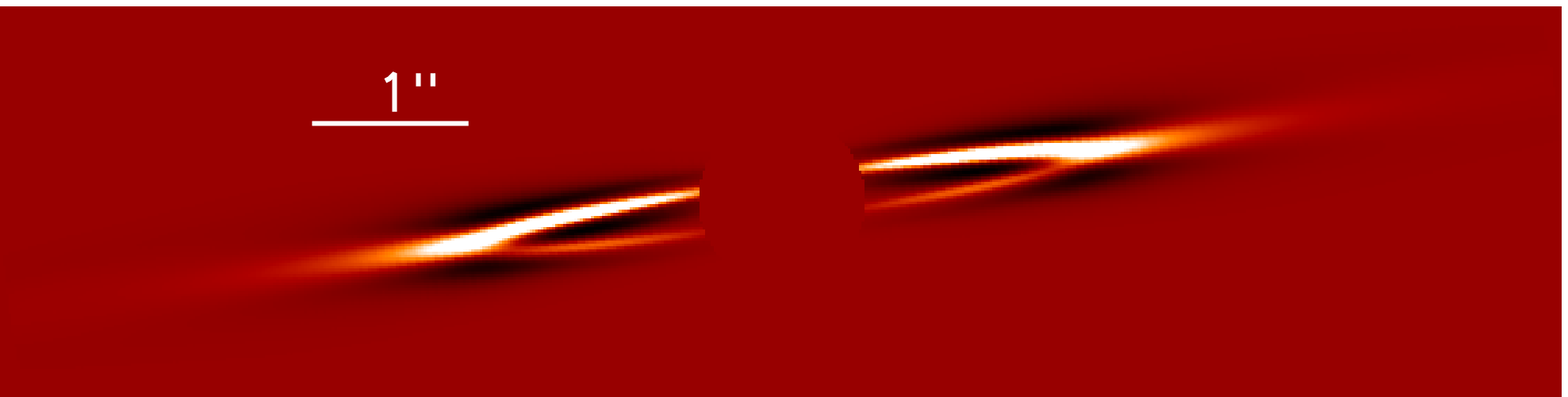}}
\caption{A simulated disk convolved with the PSF (top) and its corresponding KLIP image (bottom) assuming the observing parameters and KLIP parameters of the November 7, 2011 dataset. The model is calculated for PA=98.8$\degree$, $i=86\degree$,  $r_0=90$\,AU, $\alpha_{in}=5$, $\alpha_{out}=-5$, and $|g|=0.25$, which correspond to the best model found in Sect.~\ref{sec:fwmodel}. Colors are arbitrary and different for the two subpanels.}
\label{fig:fwmodel}
\end{figure}

In angular differential imaging, one of the main issue is the self-subtraction caused by the subtraction from the data of a reference image which may contain a fraction of the signal one attempts to detect. For point sources, like planets, this problem can be overcome with fake planets injected into the data and determined by a few parameters. However, the situation is more complex with extended sources like disks where in that case several parameters are required to determine the shape of the disk. The main purpose of the previous section was precisely to measure some geometrical parameters to restrain the parameter space in the modeling work presented below. 
A set of models was generated using the following parameters that cover a range of realistic morphologies:
\begin{itemize}
\item[-]{peak density position  ($r_0$): 85, 90, 95\,AU; }
\item[-]{power-law index of the outer density ($\alpha_{out}$): -4, -5, -6;}
\item[-]{power-law index of the inner density ($\alpha_{in}$): 2, 5, 10;}
\item[-]{anisotropic scattering factor ($|g|$): 0.00, 0.25, 0.50.}
\end{itemize}
The inclination and the PA of the models were set to 86$\degree$ and 98.8$\degree$ following the previous sections. We also set the disk aspect ratio to $h=H_0/r_0=0.01$, corresponding to a case for which the disk height is unresolved, a reasonable assumption holding for $h<0.02$ given the distance of the star.  
Contrary to previous works on inclined disks, in which the models are injected in the data at another position angle \citep{Thalmann2011, Boccaletti2012, Lagrange2012, Rodigas2012}, here we used a forward modeling approach, as proposed in \citet{Soummer2012}. The KL vectors determined in the processing of the HD\,15115 data cube are stored and reapplied on the models, which were first convolved with the PSF of the instrument. This straightforward method alleviates the problem of azimuthal noise in the image and ensures that the very same ADI bias is accounted for irrespective of the disk model. Figure \ref{fig:fwmodel} shows an example of a model image and its corresponding KLIP image calculated with the same KL vectors and the same KLIP truncation parameter as in Fig. \ref{fig:ADI_disk_big}. The disk parameters we derive below cannot be considered as definitive or very accurate since we explore only a small range of parameters with a coarse sampling. However, they help to determine the general morphology of the disk.

\begin{figure}[t]
\centerline{
\includegraphics[width=9cm]{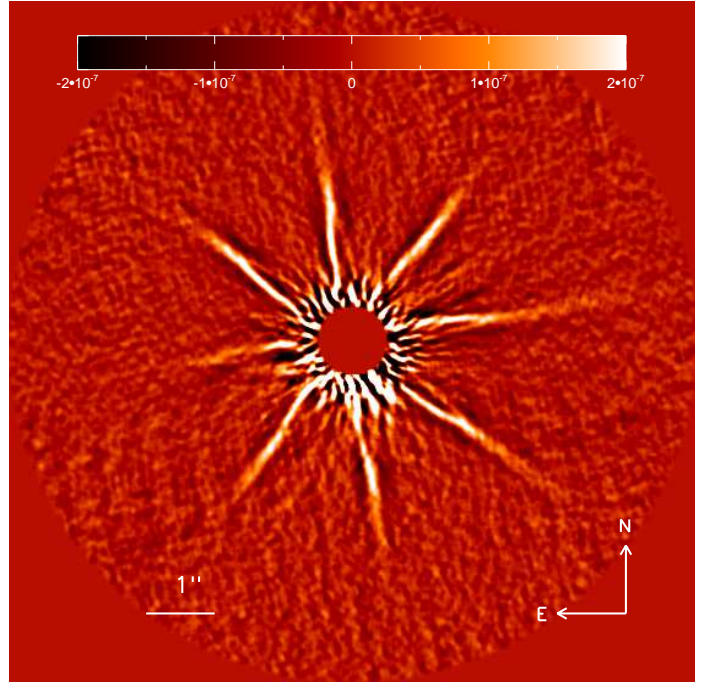}}
\caption{Same as Fig. \ref{fig:ADI_disk_big} with simulated disks injected at three different PA into the data cube and processed simultaneously. The real disk is almost horizontal. The simulated disk parameters are identical to those in Fig. \ref{fig:fwmodel}, but $\alpha_{in}=$ 2, 5, 10 (from right to left in the upper part of the image).  The images are smoothed at a scale of 1.5 resolution element. The color scale indicates the contrast with respect to the maximum intensity in the PSF image.}
\label{fig:multimodels}
\end{figure}

\begin{figure}[ht]
\centerline{
\includegraphics[width=9cm]{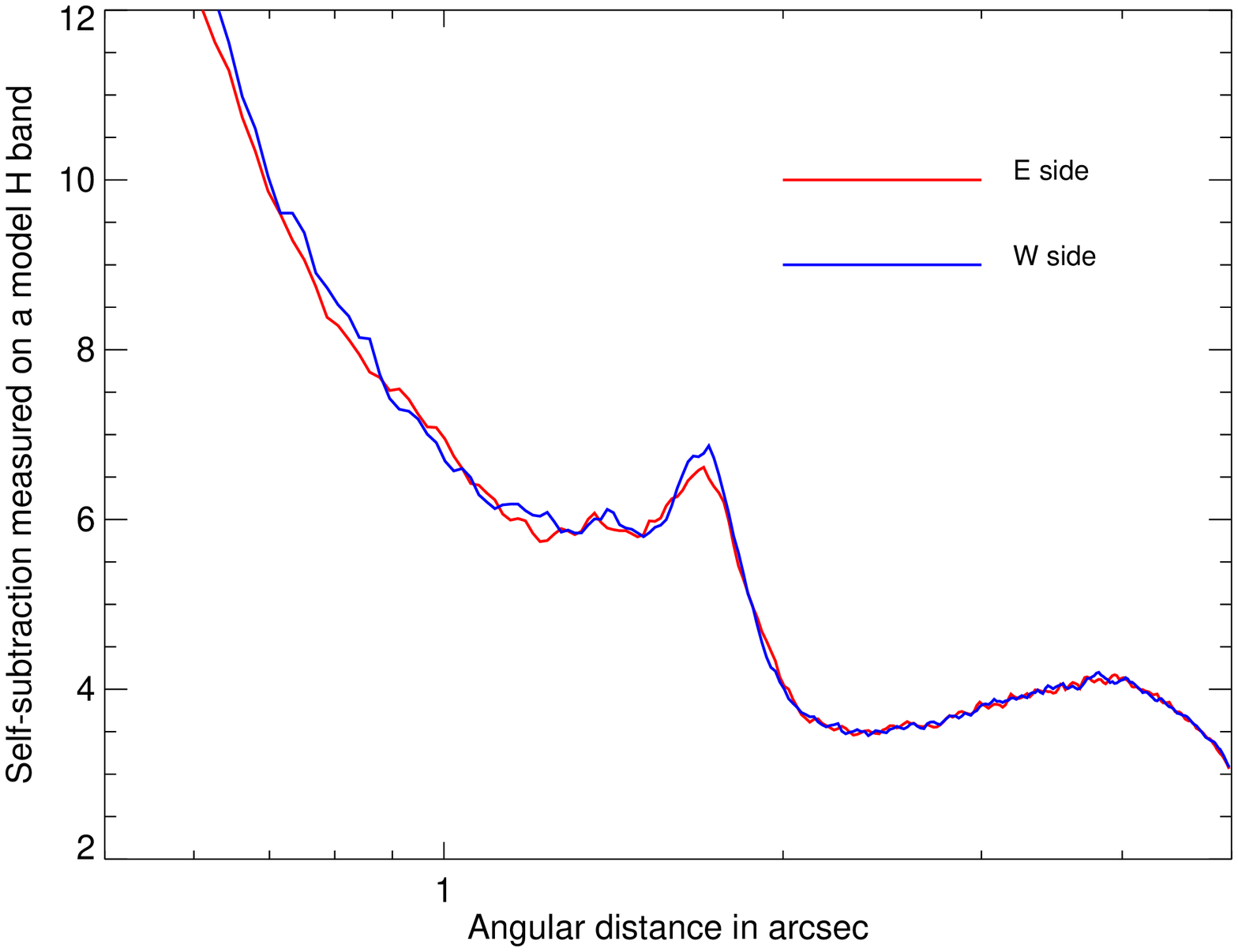}}
\centerline{
\includegraphics[width=9cm]{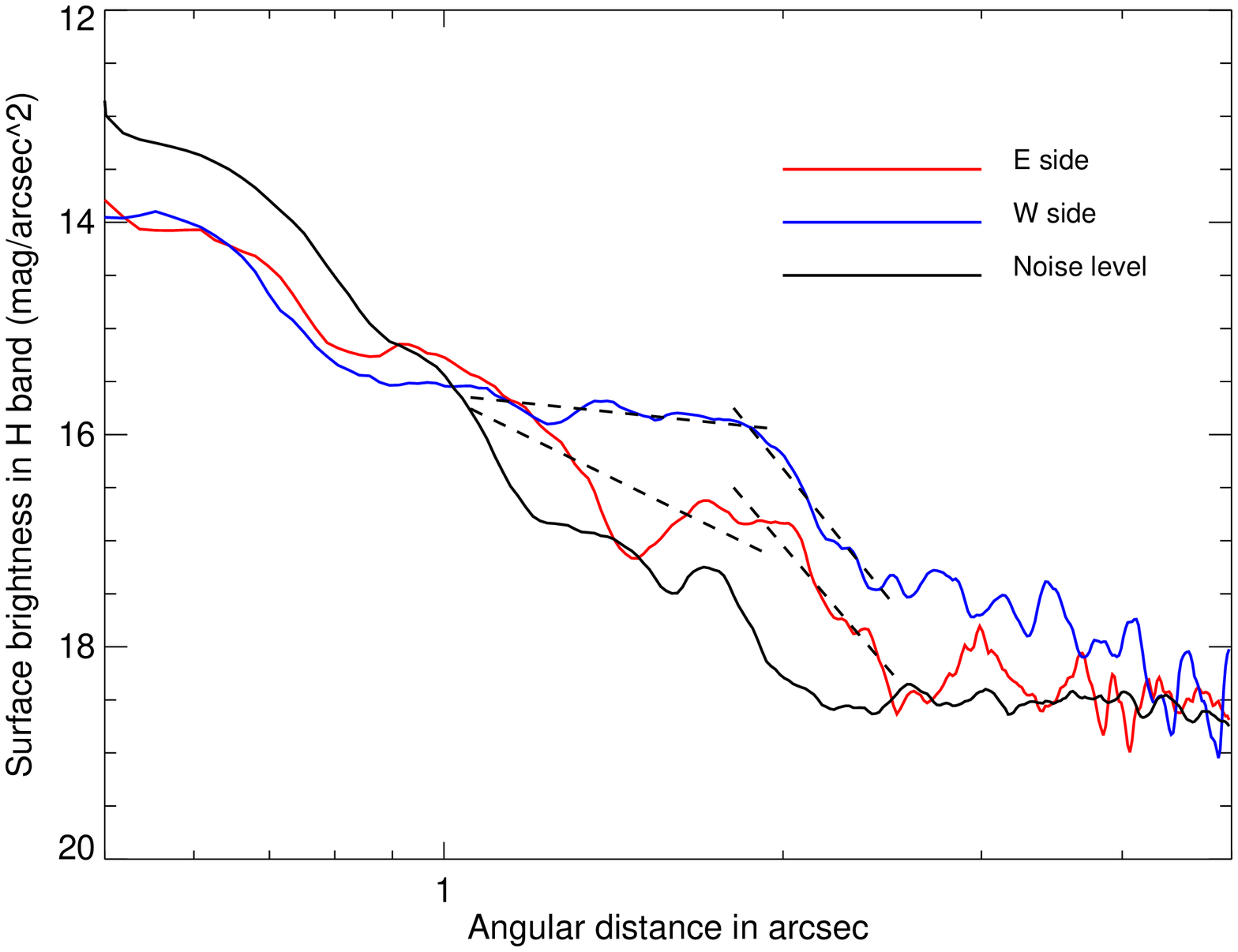}}
\caption{Intensity ratio between the simulated disk before and after processing with KLIP (top), using the model parameters given in sec. \ref{sec:fwmodel}. The SB measured in the H band image and corrected for the ADI/KLIP bias is shown in the bottom panel. Dashed lines corresponds to power laws fitted in the inner and outer regions of the disk. The black line stands for the noise level measured outside the area occupied by the disk.}
\label{fig:SB}
\end{figure}

Considering the data set of November 7, 2011, which appears of better quality, we performed a pixel-to-pixel minimization of the KLIP processed disk image and the forward KLIP models in an elliptical  region encompassing the disk (0.4" on the minor axis and extending from 1.25" to 2.0" radially). The result was not completely conclusive for all parameters but clearly favors a peak density at 90\,AU, confirming the outcome of the trace analysis, and a somehow empty cavity ($\alpha_{in}>5$), while leaving the other parameters, $\alpha_{out}$ and $|g|$, not well determined (but excluding $|g|=0$). 
Depending on whether we include feature (4) in the minimization,  $|g|$ oscillates between 0.25 (if included) and 0.50 (if not included).
To assess the remaining parameters, a subset of simulated disks was injected in the data cube at other position angles and processed simultaneously with the real disk (Fig. \ref{fig:multimodels}), so we expect azimuthal variations in that case because of the nature of the noise in the images. Visual inspection leads to set  $\alpha_{out}=-5$ and $|g|=0.25$. Therefore, the anisotropic scattering factor found with modeling is compliant with the analysis performed on the disk trace in Sect.~\ref{sec:spine}, and points to the lowest possible value of $|g|$. However, we note that these two measurements are made in two different regions of the disk, inside and outside of the peak density at 90\,AU.

To definitely rule out very different models and ensures that the central cavity was not mimicked by ADI self-subtraction, we ran the minimization on a much smaller ring (45\,AU in radius) as suggested by the SED fitting assuming blackbody grains and on a plain disk with no hole (setting $\alpha_{in}=0$). The $\chi^2$ was measured using the method described in \cite{Boccaletti2012}. In the former and latter cases, we obtained $\chi^2$ values that are $\sim2.5-5$ and $\sim1.5-3.0$ times larger than our best fit.

As a final check, we also performed the minimization on the November 22, 2011 data. The best fit yields parameters compatible with those obtained from the fit of the Nov, 7 data, that is $r_0=90$\,AU and $|g|=0.25$, while we obtained a lower value of $\alpha_{in}$ (2 instead of 5, but excluding $\alpha_{in}=0$). However, the $\chi^2$ values are much larger (by a factor of 5) than for the November 7 data, which confirms our focus on the latter data set.

\section{Photometry}
\label{sec:photom}
To measure the disk photometry we first calibrated the self-subtraction caused by the KLIP processing using the best model parameters defined above. The model convolved with the PSF is injected into an empty data cube to provide a perfect model image (using the same parallactic angle range as the real data; Fig. \ref{fig:fwmodel}, top) and a KLIP processed image  (using the same KL vectors as the data; Fig. \ref{fig:fwmodel}; bottom). The intensity of these two images is measured on concentric annuli 4 pixels (or $0.07''$) wide, centered on the star, inside a slit of 0.4", which encompasses the real disk image. The ratio of these two intensity profiles gives the photometric bias introduced by KLIP given the observing parameters (field rotation) for a particular set of disk parameters. As seen in Fig. \ref{fig:SB} (top), the shape of the self-subtraction profile is entirely correlated to the disk morphology (the peak near 2" is due to the edge of the inner cavity) and reaches a factor of 3 to 12 in the considered region (0.5" to 5"). This factor would have been different for another set of model parameters, which strengthens the importance of disk modeling for assessing photometry of disks observed in ADI. The variation of this self-subtraction according to model parameters is discussed in Appendix \ref{appendA}. 

The HD\,15115 disk intensity is measured the same way and is corrected for this factor. The star magnitude is taken from 2MASS: H=5.86 and Ks=5.82. We plot the result in Fig. \ref{fig:SB} (bottom). The noise level (black line in Fig.~\ref{fig:SB}, bottom), measured at all position angles except those where the disk is visible, indicates that the disk is detected at an angular distance larger than about 0.9-1" and closer than $\sim4"$ in the western side, and $\sim2.5-3"$ on the eastern side. The slope of the outer disk measured in the range 1.8"-2.5" is -4.98 and -5.21 for the east and west sides  (dashed lines in Fig.~\ref{fig:SB}), hence confirming the value adopted for $\alpha_{out}$. We checked with our model that for inclined disks the surface brightness (SB) slope measured in the outer disk is nearly equal to the parameter $\alpha_{out}$. 

At the edge of the ring (near 2.0"), the western side is 0.7$\sim$1.0 mag brighter than the eastern side. The inner SB profile of the west side is nearly flat (the slope is -0.44) in the 1.0"-2.0" range, while in the east side it apparently decreases with the angular separation (slope=-2.08), but the measurement is more difficult since the disk is not clearly detected at all separations. If we consider that the HD\,15115 disk is a circular ring, then the portion of the disk inside 2.0" corresponds to the same physical distances of 90\,AU, hence receiving the same amount of light from the star. Could the flat SB profile in the west be explained by the scattering function? At 1", the scattering angle is about 30$\degree$ (if the disk is edge-on) while it is 90$\degree$ at 2" (the edge of the ring). Assuming the Henyey-Greenstein phase function, we should expect a variation of about 15\% for $|g|=0.25$ between 1" and 2". Therefore, this is qualitatively consistent with the flat SB in the west. For $|g|=0.50$, the same intensity variation is instead 50\%, which could explain, again qualitatively, the slope of the eastern SB in the inner part. Hence, if the difference of slope in the inner part is real, this would favor a difference of grain properties between the east and west parts. 

In the Ks band, the SB is too noisy because of the poorer S/N than in the H band, so it is not shown here. The overall shape of the SB profile measured in the H band agrees well with those of \citet{Rodigas2012} measured in the Ks band, in particular the plateau between 1" and 2" and the steeper slope beyond 2". However, the photometric values differ by almost 2\,mag/arcsec$^2$, which cannot be explained by the star's color (H-Ks=0.04). As the photometric calibration was extrapolated from the PSF halo of a much fainter star, we assume that these LBT data do not have a clean PSF calibration for a proper comparison with our result.
On the contrary, the HST photometry measured by \citet{Debes2008} on the west side at 2" reaches $\sim$15.6\,mag/arcsec$^2$ in the F110W filter (1.1$\mu m$), while we find $\sim$16.1\,mag/arcsec$^2$. These two values agree more closely, and are consistent with the blue color reported by these authors. Finally, \citet{Kalas2007} reported at the same angular separation a SB of $\sim$16.4\,mag/arcsec$^2$ in the H band from Keck observations that are again in good agreement with our work. However, they did not detect a SB flattening inwards $2''$ and concluded that the inner dust depletion should reside within 40\,AU, while we do see such a feature in our image in the $\sim 45-90$ AU region \citep[also proposed by][]{Schneider2014}.
The near-IR colors of the HD\,15115 disk should be revisited in light of the latest observations collected with the same angular resolution from Keck, LBT, and Gemini South.

\section{Summary and discussion}
\label{sec:ccl}

In this article, we presented H, Ks, and CH4 band images of the debris disk around HD\,15115 from archival data sets obtained with NICI in 2009 and 2011.
Several ADI algorithms were used to process the data and allow the disk to be detected at all epochs and all filters.

The general characteristics of the disk inferred from previous studies, its nearly edge-on geometry, the east-west brightness asymmetry, and the bow-like shape are confirmed. 
However, the NICI data allow us to detect the ansae on both sides and to infer the presence of an inner cavity with higher confidence than ever before. The disk around HD\,15115 has, very likely, a ring-like shape as does the disk around HD\,32297, for instance. However, only the upper half of the disk is firmly detected. The lower part, labeled (4) in our images, remains difficult to confirm since it does not show up in all datasets. The ansae of this ring are located at a projected radius of 1.99" ($\simeq 90$\,AU) which differs by a factor of 2 from the cavity size derived by SED fitting. While the brightness asymmetry is  obvious at large separations beyond the peak of density at 90\,AU, the ring is in fact very symmetrical on both sides of the star and does not feature any significant offsets with respect to the central star. Given a large inclination (86$\degree$) it would be almost impossible to measure an offset along the line of sight. 

We were able, based on a disk geometrical model, to correlate the misalignment and the elevation from the midplane of the two sides of the disk with the anisotropic scattering factor $|g|$. Therefore, the mean PA= $98.8\pm0.4 \degree$ of the two disk sides defines the midplane and we measured an elevation in the 2-3" region of $0.03\pm0.01"$ corresponding to $0.2<|g|<0.5$ in the outer disk. 

The forward modeling approach enables us to constrain some of the disk parameters as well as to correct the photometry from ADI biases. We found that the inner cavity is relatively empty ($\alpha_{in}=5$) as suggested by the ring-like shape in Fig. \ref{fig:ADI_disk_big}, while the anisotropic scattering factor is close to a value of $|g|=0.25$. As we performed several assessments (geometrical or photometrical) of $|g|$ in different parts of the disk, the dispersion of the values that we found can just be related to a variation of grain properties between the nearest regions at the edge of the cavity at 90\,AU and the regions farther out populated by grains blown away by radiation pressure.

The main result in this paper is the symmetry of the inner ring, which contrasts with the east-west brightness asymmetry of both the ring and the regions outside the ring. Interaction with the local ISM has been proposed as a plausible explanation to account for the brightness asymmetry \citep{Rodigas2012}. A thorough modeling remains to be done to understand this effect, but we can provide some hypotheses for an alternative or additional mechanism. The east-west brightness asymmetry cannot be explained by a geometrical offset of the ring as it can be in the case of the pericenter glow \citep{Wyatt1999}. Instead, it can be related to a local increase in the scattering cross section of the dust grains and/or a local density enhancement of small grains, both of which could point toward an increase in the collisional activity at the place where the ring appears the brightest. For instance, a recent collision could explain a variation of the grain size and/or property distributions \citep{Jackson2014}, an explanation which was also proposed for the clump in the $\beta$ Pic disk observed by \cite{Telesco2005} and \cite{Dent2014} at mid-IR and submillimeter wavelengths, respectively. This could also provide theoretical support to the wavelength dependence of the brightness asymmetry seen by \citet{Debes2008} who suggested the presence of several populations of grains with various properties. Other debris disks, like HD 61005 \citep{Buenzli2010}, also feature a similar brightness asymmetry combined with a ring-like geometry. There is still progress to be made to understand young planetary systems like the one around HD15115. In this respect, the upcoming generation of high-contrast instruments will be major contributors. 

\begin{appendix}
\label{append}
\section{Self-subtraction}
\label{appendA}
\begin{figure}[ht]
\centerline{
\includegraphics[width=9cm]{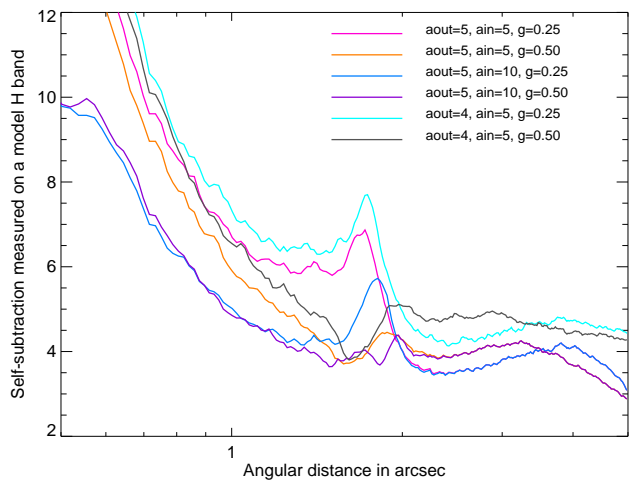}}
\caption{Intensity ratio between the simulated disk before and after processing with KLIP (top), using the model parameters given in Sect. \ref{sec:fwmodel}.}
\label{fig:selfsub}
\end{figure}

We calculated the self-subtraction of the disk for various model parameters $\alpha_{out}$, $\alpha_{in}$, and $|g|$ while other parameters are fixed: $i=86\degree$, $r_0$=90\,AU. The amount of attenuation is shown in Fig. \ref{fig:selfsub}.
We observe important variations (20\% to 30\% ) inside the cavity at separations shorter than 1"-1.5" (but the disk remains undetected inside 1") and outside the cavity ($>2''$). The various values of $\alpha_{in}$ and $\alpha_{out}$ affect the level of the self-subtraction but do not change the slope of this profile. On the contrary, the most important effects are observed near the edge of the cavity between 1.5" and 2" ($r_0$=90\,AU= 2"). In this case, $\alpha_{in}$  and $|g|$ are the dominant contributors and can induce a variation of almost a factor of 2 at this angular separation. The peak at the edge of the cavity can be shifted in radial position. Overall, this translates into a potential impact of 0.25 to 0.75 mag on the SB profile. Still, this does not explain the discrepancy with \citet{Rodigas2012} on the disk photometry, but definitely calls for a careful analysis of the disk geometry in order to recover the exact photometry.

\end{appendix}

\begin{acknowledgements}
J. Mazoyer is grateful to the Centre National d'Etudes Spatiales (CNES, Toulouse, France) and Astrium (Toulouse, France) for supporting his PhD fellowship. We would also like to thank the referee for a very detailed and useful report which helped us to improve the manuscript.
 \end{acknowledgements}

\bibliographystyle{aa} 
\bibliography{hd15115_bib}   

\end{document}